\documentclass{article}
\usepackage[utf8]{inputenc}
\usepackage{authblk}

\usepackage{amsmath}
\usepackage{bbold}
\usepackage{braket}
\usepackage{graphicx}
\usepackage{xcolor}

\title{Parallel Gate Operations Fidelity in a Linear Array of Flip-Flop Qubits}

\author{Davide Rei}
\author{Elena Ferraro}
\author{Marco De Michielis}
\affil{CNR-IMM Agrate Unit, Via C. Olivetti 2, 20864 Agrate Brianza (MB), Italy}

\begin{document}

\maketitle
\begin{abstract}
Quantum computers based on silicon are promising candidates for long term universal quantum computation due to the long coherence times of electron and nuclear spin states. Furthermore, the continuous progress of micro- and nano- electronics, also related to the scaling of Metal - Oxide - Semiconductor (MOS) systems, makes possible to control the displacement of single dopants thus suggesting their exploitation as qubit holders. Flip-flop qubit is a donor based qubit (DQ) where interactions between qubits are achievable for distance up to several hundred nanometers. In this work, a linear array of flip-flop qubits is considered and the unwanted mutual qubit interactions due to the simultaneous application of two one-qubit and two two-qubit gates are included in the quantum gate simulations. In particular, by studying the parallel execution of couples of one-qubit gates, namely $R_z(-\frac{\pi}{2})$ and $R_x(-\frac{\pi}{2})$, and of couples of two-qubit gate, i.e. $\sqrt{iSWAP}$, a safe inter-qubit distance is found where unwanted qubit interactions are negligible thus leading to parallel gates fidelity up to 99.9\%. 
\end{abstract}

\section{Introduction}

Donor atoms in silicon represent a well-known system to host qubits due to their potential for scalability and affinity with MOS technology, the transistor reference technology. This kind of qubit was proposed by Kane in 1998 where the nuclear spin states of a phosphorus donor atom into a silicon bulk define the qubit \cite{Kane-1998}. The physical properties of these elements are well known because are commonly used in semiconductor industry making these materials one of the first choices to realize a solid state quantum computer. 

Moreover, silicon offers a long coherence time of electron and nuclear spin states and, in particular, its isotope $^{28}$Si is spin-free, so the interactions between the spins of donor and silicon bulk, which compromise the coherence states of qubits system, are avoided \cite{Gordon-1958} \cite{Feher-1959} \cite{Tyryshkin-2012} \cite{Steger-2012} \cite{Saeedi-2013} \cite{Morello-2020}.

The realization of a two-qubit operation, which requires an interaction between qubits, is needed to implement a quantum algorithm. The interaction exploited in the Kane's qubit is the exchange interaction, that acting on a short range imposes an accurate donor placement. This circumstance represents one of the main issue related to the realization of this architecture.

Recently, the flip-flop qubit, a particular type of DQ, has been studied to overcome this limitation \cite{Tosi-2017}, \cite{Tosi-2018}, \cite{Calderon-2021}, \cite{Ferraro-2021}. It is constituted by a phosphorus donor atom embedded in a $^{28}$Si substrate displaced at a distance \textit{d} from a SiO$_2$ interface. At the top, a metal gate generates an electric field $E_z$ to control the donor-bound electron position between the nucleus and the interface with the oxide. In this way, not only it is possible to define a qubit but also an electric dipole is created by the negatively charged electron and the positively charged donor nucleus. Taking advantage of the dipole-dipole interaction, it enables the coupling between two qubits up to distances of an order of magnitude higher than the Kane’s qubit, in the 100-500 nm range. The feature related to the long-distance interaction between two qubits relaxes the fabrication accuracy on metal gates and $^{31}$P donors position needed to scale up the system. The interconnection between qubits is eased and the formation of a logical qubit, that is a system of more physical qubits which state is used to encode the state of a qubit, can be performed to protect the information by exploiting quantum error correction (QEC) codes \cite{Steane-1996}. Depending on the QEC codes, some logical gates can be transversal, meaning that the operation on the logical qubit is simply obtained by applying the operation to each physical qubit, i.e. in a bit-wise fashion \cite{Nielsen-2000}. In order to take fully advantage of transversal gates, the effects of unwanted interactions between flip-flop qubits manipulated in parallel need to be studied.

A universal quantum computer requires indeed the capability of performing single-qubit gates and two-qubit entangling operations in parallel \cite{Yousefjani-2021} \cite{Figgatt-2019}. Study about quantum gate parallelism have been carried out on superconducting circuits \cite{Song-2017}, chains of atoms \cite{Olsacher-2020} \cite{Landsman-2019} \cite{Figgatt-2019} \cite{Levine-2019} and spins \cite{Yousefjani-2021}. In this work we address this kind of study on a linear array of flip-flop qubit.

The paper is organized as follows. In section \ref{Sec:System} the one-qubit system Hamiltonian and the definition of the flip-flop qubit states are introduced. Then, in sections \ref{Sec:ParallelOneQubitOp} and \ref{Sec:ParallelTwoQubitOp}, studying the states evolution during the interaction with the external control electric field, different one-qubit gates, namely $R_z(-\frac{\pi}{2})$, $R_x(-\frac{\pi}{2})$ and a two-qubit operation, i.e. the entangling $\sqrt{iSWAP}$ gate, are presented and the study of the gate fidelity for these operations applied in parallel between two qubits and two couples of qubits is investigated. Finally, in section \ref{Sec:Conclusion} the conclusions of this work are provided.

\section{Flip-flop qubit}
\label{Sec:System}

The flip-flop qubit is described in the eight-dimensional Hilbert space that takes into account the spin states of the donor electron (nucleus) $\{ \ket{\downarrow}$; $\ket{\uparrow} \}$ ($\{ \ket{\Downarrow}$; $\ket{\Uparrow} \}$) and the orbital degree of freedoms $\{ \ket{g}$; $\ket{e} \}$. The energy difference between the electron ground $\ket{g}$ and excited $\ket{e}$ states is given by \cite{Tosi-2017}, \cite{Simon-2020}
\begin{equation}
\epsilon_0
=
\sqrt{V_t^2 + \left( \frac{d e ( E_z - E_z^0 ) }{h} \right)} ,
\end{equation}
where $E_z - E_z^0 \equiv \Delta E_z$ is the difference between the vertical electric field $E_z$ applied by the gate and its value $E_z^0$ at the ionization point, where the electron is shared halfway between donor and interface. $V_t$ is the tunnel coupling between the donor and the interface potential wells, $e$ is the elementary charge, $h$ is the Planck constant and $d$ is the distance between the nucleus and the interface.


The Hamiltonian $\hat{H}$ describing the flip-flop qubit is composed by the Zeeman part $\hat{H}_{B}$, the hyperfine coupling term $\hat{H}_{A}$ and the orbital part $\hat{H}_{Orb}$ \cite{Tosi-2017}:

\begin{equation}
\label{eq:Hff}
\hat{H} = \hat{H}_{B_0} + \hat{H}_A + \hat{H}_{Orb} .
\end{equation}

Each term can be written as a function of the Pauli matrices

\begin{equation}
\hat{\sigma}_z = \ket{g} \bra{g} - \ket{e} \bra{e} ,
\end{equation}
\begin{equation}
\hat{\sigma}_x = \ket{g} \bra{e} + \ket{e} \bra{g}
\end{equation}
and the electron (nuclear) spin operators \textbf{S} (\textbf{I}), with $\hat{z}$ component $\hat{S}_z$ ($\hat{I}_z$).

The first two terms $\hat{H}_{B}$ and $\hat{H}_{A}$ which have the following expressions:

\begin{equation}
\label{Eq:HB0}
\hat{H}_{B_0} = \gamma_e B_0 \left[ \hat{\mathbb{1}} + \left( \frac{\hat{\mathbb{1}}}{2} + \frac{ d e \Delta E_z }{2 h \epsilon_0} \hat{\sigma}_z + \frac{V_t}{2 \epsilon_0} \hat{\sigma}_x \right) \Delta_{\gamma} \right] \hat{S}_z - \gamma_n B_0 \hat{I}_z ,
\end{equation}

\begin{equation}
\hat{H}_A = A \left( \frac{\hat{\mathbb{1}}}{2} - \frac{ d e \Delta E_z }{2 h \epsilon_0} \hat{\sigma}_z - \frac{V_t}{2 \epsilon_0} \hat{\sigma}_x \right) \textbf{S} \cdot \textbf{I} ,
\end{equation}
describe the Zeeman splitting caused by a constant magnetic field $B_0$ and the hyperfine interaction, respectively. In particular, in Equation \ref{Eq:HB0}, $\Delta \gamma$ takes into account the variation of the electron gyromagnetic ratio $\gamma_e$ between the nucleus (27.97 GHz/T) and the interface, while $\gamma_n = 17.23$ MHz/T is the constant nuclear gyromagnetic ratio. The hyperfine coupling $A$ is a function of the vertical electric field $E_z$ applied by the gate and, in order to obtain its functional form, the results reported in ref. \cite{Tosi-2017} are fitted with the function $A_0/\left( 1 + \text{exp} \left( c \Delta E_z \right) \right)$, where $A_0 = 117$ MHz is the bulk value of $A$, obtaining the fit parameter $c = 5.174 \cdot 10^{-4}$ m/V \cite{Ferraro-2021}. Finally, the operator $\hat{\mathbb{1}}$ is the identity matrix.

The orbital part $\hat{H}_{Orb}$, which gives a treatment of the electron position between the interface and the donor as a two level system allowing a full quantum mechanical description of the system, is given by

\begin{equation}
\hat{H}_{Orb} = - \frac{\epsilon_0}{2} \hat{\sigma}_z - \frac{ d e E_{ac}(t) \text{cos} ( \omega_E t +\phi) }{2 h} \left( \frac{d e \Delta E_z }{h \epsilon_0} \hat{\sigma}_z + \frac{V_t}{\epsilon_0} \hat{\sigma}_x \right) ,
\end{equation}
where $E_{ac}(t)$ is the time dependent amplitude of an oscillating electric field with pulsation $\omega_E$ and phase $\phi$. The qubit states are defined as the tensor product between the electron ground state and the flip-flop antiparallel states, i.e. $\ket{0} \; \equiv \; \ket{g \downarrow \Uparrow}$ and $\ket{1} \; \equiv \; \ket{g \uparrow \Downarrow}$.

For the study of the quantum operations carried out in the next sections the parameters reported in ref. \cite{Tosi-2017} are used, i.e. $B_0 = 0.4$ T, $\Delta \gamma = -0.002$ and $d = 15$ nm. 

\section{Parallel one-qubit gates}

\label{Sec:ParallelOneQubitOp}

\begin{figure}[h!]
    \centering
    \includegraphics[width=0.5\textwidth]{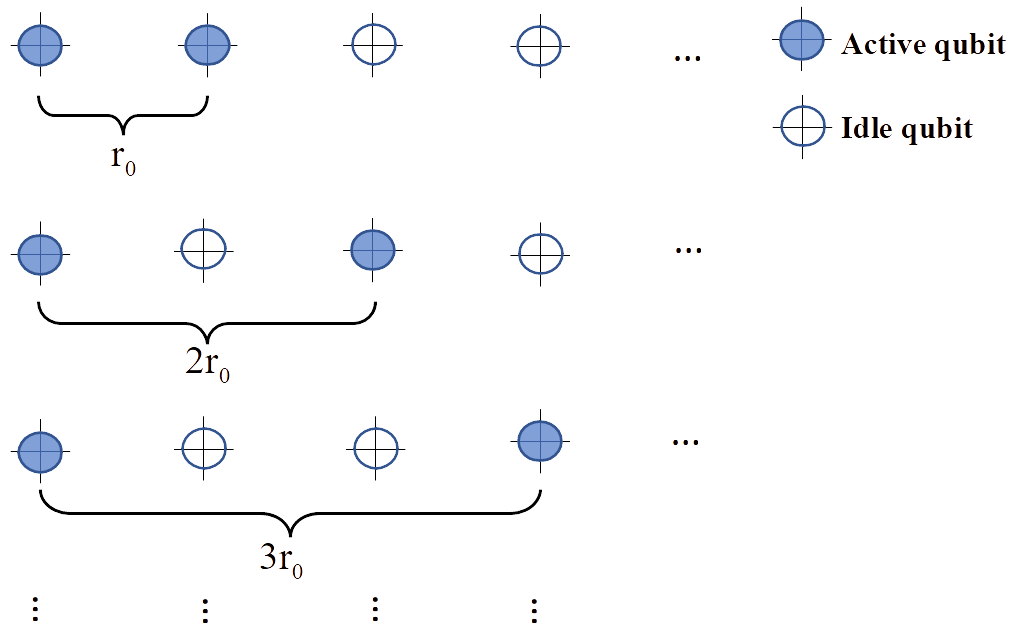}
    \caption{Scheme of a linear array of flip-flop qubits in three different example configurations. Each qubit is positioned at a distance $r_0$ from the adjacent one and two parallel one-qubit operations are executed on qubits displaced by a distance $r$, multiple of $r_{0}$. 
    Qubits between active ones are assumed in idle.}
    \label{Fig:ParallelOneQubitOp}
\end{figure}

In this section the focus is on the effects of the unwanted interactions between two qubits in a linear array operated with one-qubit gates in parallel. The two qubits are displaced at an inter-qubit distance $r$, with $r$ an integer multiple of the nearest-neighbor qubit distance $r_{0}$ as shown in \textbf{Figure \ref{Fig:ParallelOneQubitOp}}. $r_{0}$ is the reference spacing and it is set to 180 nm that is the value used to extract the control sequence of the two-qubit gate, i.e. $\sqrt{iSWAP}$, in this study \cite{Ferraro-2021}.

The interaction between qubits is mediated by the long range dipole-dipole interaction between the two electric dipoles at the qubit sites electrically induced by the displacement of the electron of each donor atom toward the interface \cite{Tosi-2017}. 

Assuming identical flip-flop qubits with indexes $i$ and $j$, the Hamiltonian of the two-qubit system $\hat{H}_2^{ij}$ is the sum of the two single-qubit Hamiltonians $\hat{H}$ (Equation \ref{eq:Hff}) and the interaction term

\begin{equation}
\label{eq:H2ff}
\hat{H}_2^{ij}=\hat{H}^i\otimes\hat{\mathbb{1}}+\hat{\mathbb{1}}\otimes\hat{H}^j+\hat{H}_{int}^{ij} ,
\end{equation}
where
\begin{equation}
\label{eq:Vdip0}
\hat{H}_{int}^{ij}=\frac{1}{4\pi\epsilon_0\epsilon_r r^3} \left[\textbf{p}_i\cdot\textbf{p}_j-\frac{3(\textbf{p}_i\cdot\textbf{r})(\textbf{p}_j\cdot\textbf{r})}{r^2}\right]
\end{equation}
is the dipole-dipole interaction. Here, $\epsilon_0$ is the vacuum permittivity, $\epsilon_r$ is the material dielectric constant (equals to 11.7 for silicon), $\textbf{r}$ is the vector distance between the two qubits and $\textbf{p}_{i(j)}=\frac{ed}{2}\left(\hat{\mathbb{1}}_{i(j)}+\hat{\sigma}_{z,i(j)}^{id} \right)$ is the dipole operator of the qubit to whom is associated the position operator 
\begin{equation}
\hat{\sigma}_z^{id} \; = \; \frac{d \, e \, \Delta E_z }{h \, \epsilon_0} \hat{\sigma}_z \; + \; \frac{V_t}{\epsilon_0} \hat{\sigma}_x ,
\end{equation}
whose eigenstates $\ket{i}$ and $\ket{d}$ indicate if the electron is localized near the interface or the donor, respectively.

In the following, the effects of the unwanted interactions between qubits on the gate infidelity when two one-qubit gates are applied in parallel are considered. The study is performed taking into account two active qubits separated by none, one, two and three idling qubits. The qubit idling state is obtained by keeping the electron near to the nucleus in order to switch off its electric dipole. In this way the idle qubits can not interfere via dipole-dipole interaction with the active qubits.

The operations studied in this section are the $R_z(-\frac{\pi}{2})$ and $R_x(-\frac{\pi}{2})$ rotations which will be applied individually on each qubit. Like in ref. \cite{Ferraro-2021}, the entanglement fidelity F \cite{Nielsen-2000} is calculated for each gate when the 1/f noise model on the electric field $\Delta E_z$ is considered\cite{Paladino-2014} \cite{Epstein-2014} \cite{Yang-2016} \cite{Zhang-2017} \cite{Yang-2019PRA} \cite{Ferraro-2020-sr}.

\subsection{Parallel $R_{z}$ gates}

A rotation around the $\hat{z}$-axis of the Bloch sphere can be obtained by exploiting the phase accumulation between the two qubit states that is generated during the interaction of the system with an external electric field. To do this, a DC electric field $\Delta E_z (t)$ is swept from an idling value $\Delta E_{idle}$, where the electron is confined at the interface, to an intermediate value $\Delta E_{int}$ in a time $\tau_1$. Then, a clock transition value for the electric field $\Delta E_{ct}$, where the dephasing rate is minimum \cite{Tosi-2017}, is reached after a time $\tau_2$ and, after a time $T$ which sets the angle of rotation, the electric field is reset back to the idling value, following backwards the previous sequence steps. The parameters used to set a $-\pi/2$ rotation with an adiabaticity value of $K \simeq 20$ are calculated following ref. \cite{Tosi-2017} and are shown in Table \ref{tab:Rzm05pi}.

\begin{table}[htbp!]
 \begin{center}
 \caption{Single $R_z(-\frac{\pi}{2})$ gate parameters.}
  \begin{tabular}[htbp]{@{}llllllll@{}}
    \hline
   $V_t$ & $\Delta E_{idle}$ & $\Delta E_{int}$ & $\Delta E_{ct}$ & $\tau_1$ & $\tau_2$ & T &  K \\
    \text{[GHz]} &\text{[V/m]} & \text{[V/m]} & \text{[V/m]} & \text{[ns]} & \text{[ns]} & \text{[ns]} &  \\
    \hline
    11.29 &10000 & 1300 & 290 & 2 & 16 & 0.08 &  $\simeq$ 20 \\
    \hline
  \end{tabular}
      \label{tab:Rzm05pi}
   \end{center}
\end{table}

\begin{figure}[htbp!]
    \centering
    \includegraphics[width=0.8\textwidth]{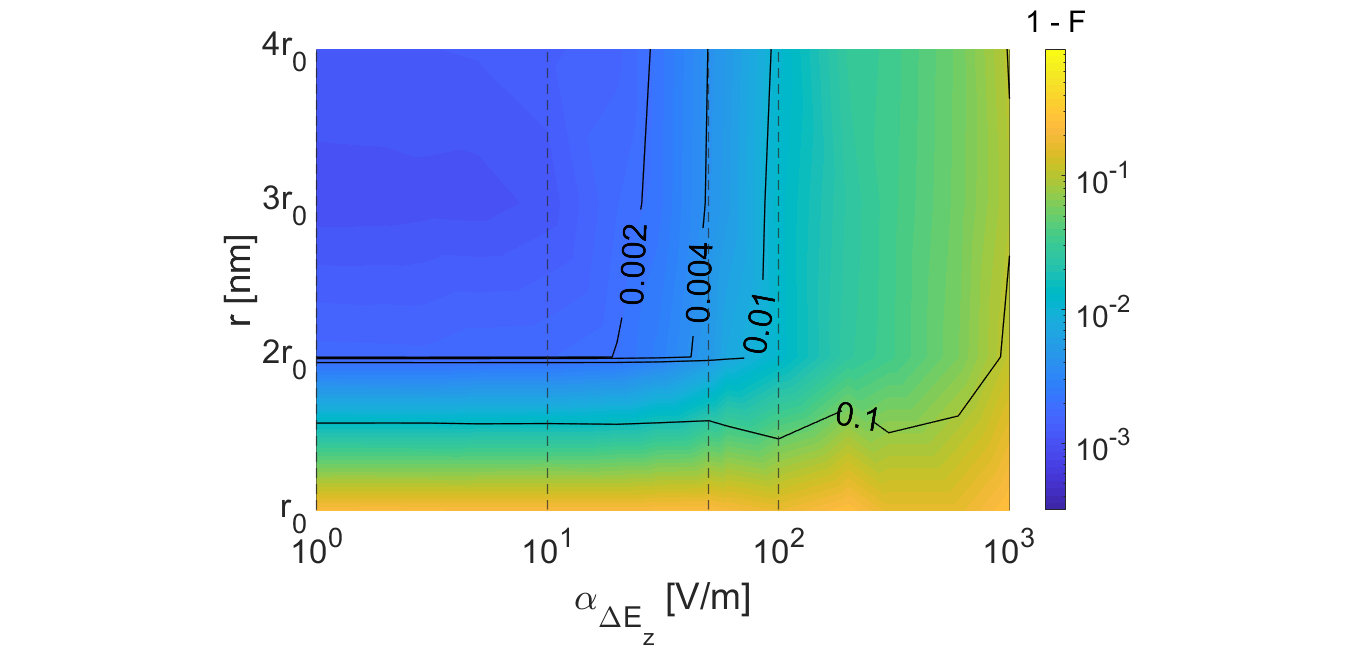}
    \caption{Entanglement infidelity for parallel one-qubit operation $R_z ( -\frac{\pi}{2} ) \otimes R_z ( -\frac{\pi}{2} )$ as a function of noise amplitude $\alpha_{\Delta E_z}$ and $r$. Infidelity is mostly deteriorated in the region close to $r = r_0$ due to a stronger dipole-dipole interaction and where $\alpha_{\Delta E_z}$ is higher. The dashed vertical lines highlight the $\alpha_{\Delta E_z}$ values investigated in the next Figure.}
    \label{Fig:Inf-alpha-rcouple_ParRz-0.5pi}
\end{figure}

In order to quantify the combined effects of the inter-qubit distance $r$ and of the 1/f noise amplitude $\alpha_{\Delta E_z}$ on two parallel $R_z(-\frac{\pi}{2})$ gates, the infidelities 1-F are presented in the equi-infidelity graph of \textbf{Figure \ref{Fig:Inf-alpha-rcouple_ParRz-0.5pi}} for $r$ ranging from $r_0$ to 4$r_0$ and for an $\alpha_{\Delta E_z}$ range spaced between 1 and 1000 V/m. 

A lower infidelity is generally obtained for $r > r_0$ and for smaller $\alpha_{\Delta E_z}$.
The worst infidelities are obtained for $r = r_0$ and for high $\alpha_{\Delta E_z}$. When $r = r_0$, the high values of 1-F are due to the long-range inter-qubit dipole-dipole interaction that is sufficiently strong to compromise the nearby parallel operation, while for $r \geq 2 r_0$, the inter-qubit distance is enough to reduce the qubits interaction leaving practically unaffected the parallel operations. 
When $\alpha_{\Delta E_z}$ is increased, 1-F increases up to exceed $10^{-2}$ when $\alpha_{\Delta E_z} = 100$ V/m for $r \geq 2 r_0$.

\begin{figure}[htbp!]
    \centering
    \includegraphics[width=0.8\textwidth]{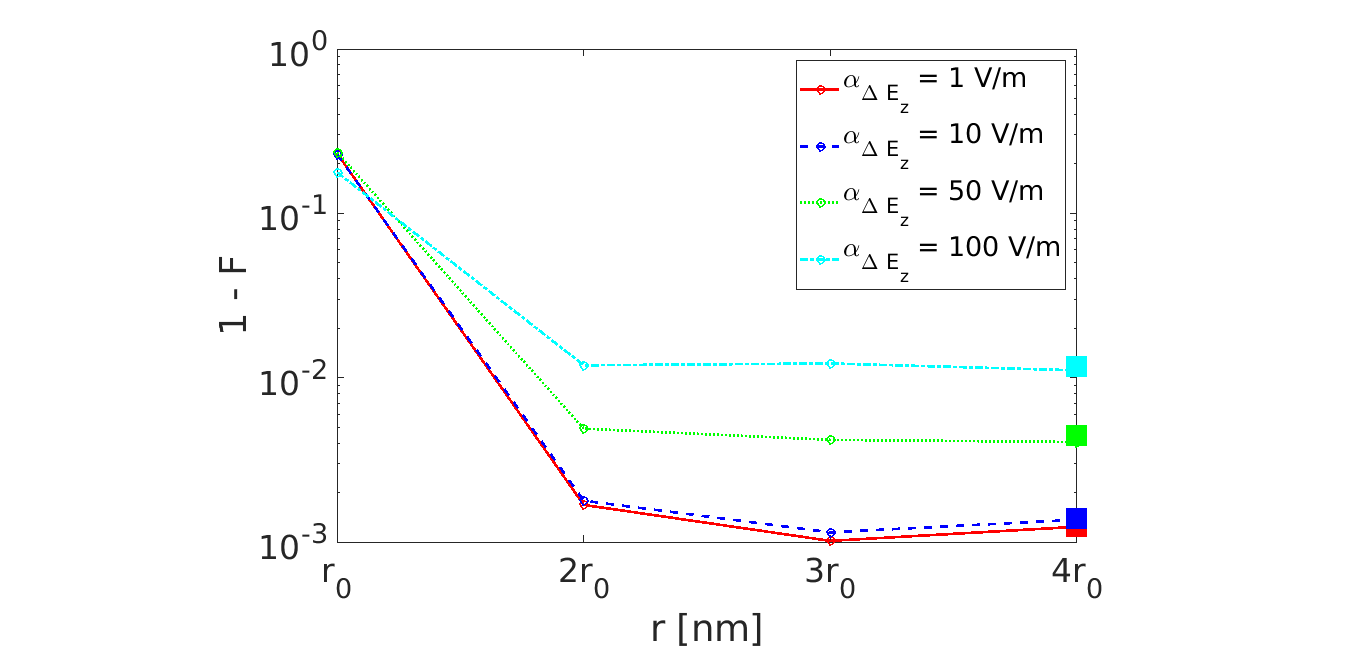}
    \caption{Entanglement infidelity for parallel one-qubit operation $R_z ( -\frac{\pi}{2} ) \otimes R_z ( -\frac{\pi}{2} )$ as a function of $r$ for different noise amplitudes $\alpha_{\Delta E_z}$. When $r \geq 2 r_0$, the dipole-dipole interaction strength is essentially negligible thus the 1-F curves are flat. The squares represent the infidelity of two non-interacting qubits for each value of $\alpha_{\Delta E_z}$.}
    \label{Fig:Inf-r_ParRz-0.5pi}
\end{figure}

In \textbf{Figure \ref{Fig:Inf-r_ParRz-0.5pi}} the effect of the qubits distance on the infidelities of two parallel $R_z(-\frac{\pi}{2})$ gates is shown for different significative values of $\alpha_{\Delta E_z}$ equal to 1, 10, 50 and 100 V/m. The long range dipole-dipole interaction strength causes the 1-F curves maximum value at $r = r_0$ while the infidelities are approximately flat due a reduced interaction for $r \geq 2 r_0$. In this last region the main contribution to the infidelity deterioration is due to $\alpha_{\Delta E_z}$. We point out that the fidelity reaches 99.9\% in correspondence to $\alpha_{\Delta E_z}\leq$ 10 V/m, 99.6\% for $\alpha_{\Delta E_z}=$ 50 V/m and 99\% for $\alpha_{\Delta E_z}=$ 100 V/m. For comparison, the squares highlighting the infidelity of two non-interacting qubits for each value of $\alpha_{\Delta E_z}$ in the corresponding colour are added.

\subsection{Parallel $R_{x}$ gates}

Unlike $R_{z}$, a $R_{x}$ gate needs the addition to the DC electric field of an AC electric field

\begin{equation}
E_a (t)
=
E_{ac} (t) \text{cos} \left( 2 \pi\epsilon_{ff} t \right) ,
\end{equation}
in resonance with the flip-flop qubit transition frequency at $\Delta E_{ct}$, where $E_{ac} (t)$ is the electric field amplitude with a triangular envelope which drives the rotation around the $\hat{x}(\hat{y})$-axis, $\epsilon_{ff}$ is the flip-flop qubit transition frequency associated to the qubit states energy difference and the electric field phase $\phi=0$. The oscillating field is summed to the DC component after a time $T_{E_{ac}}^{\text{Start}}$ for a duration $T_{E_{ac}}^{\text{ON}}$. The parameters shown in Table \ref{tab:Rxm05pi} are used to obtain a $-\pi/2$ rotation with an adiabaticity value of $K \simeq 20$.

\begin{table}[htbp!]
  \begin{center}
    \caption{Single $R_x \left( - \frac{\pi}{2} \right)$ gate parameters.}
    \begin{tabular}{@{}lllllllllll@{}}
    \hline
      $V_t$ & $\Delta E_{idle}$ & $\Delta E_{int}$ & $\Delta E_{ct}$ & $\tau_1$ & $\tau_2$ & T &$\max(E_{ac} (t))$ & $T_{E_{ac}}^{\text{Start}}$ & $T_{E_{ac}}^{\text{ON}}$ &   K \\
      \text{[GHz]} &\text{[V/m]} & \text{[V/m]} & \text{[V/m]}& \text{[ns]} & \text{[ns]} & \text{[ns]} & \text{[V/m]} & \text{[ns]} & \text{[ns]} &  \text{} \\
      \hline
      11.5 &10000 & 1300 & 0 & 2 & 4 & 90.5 & 180 & 25 & 40 &  $\simeq$ 20 \\
      \hline
    \end{tabular}
    \label{tab:Rxm05pi}
  \end{center}
\end{table}

\begin{figure}[htbp!]
    \centering
    \includegraphics[width=0.8\textwidth]{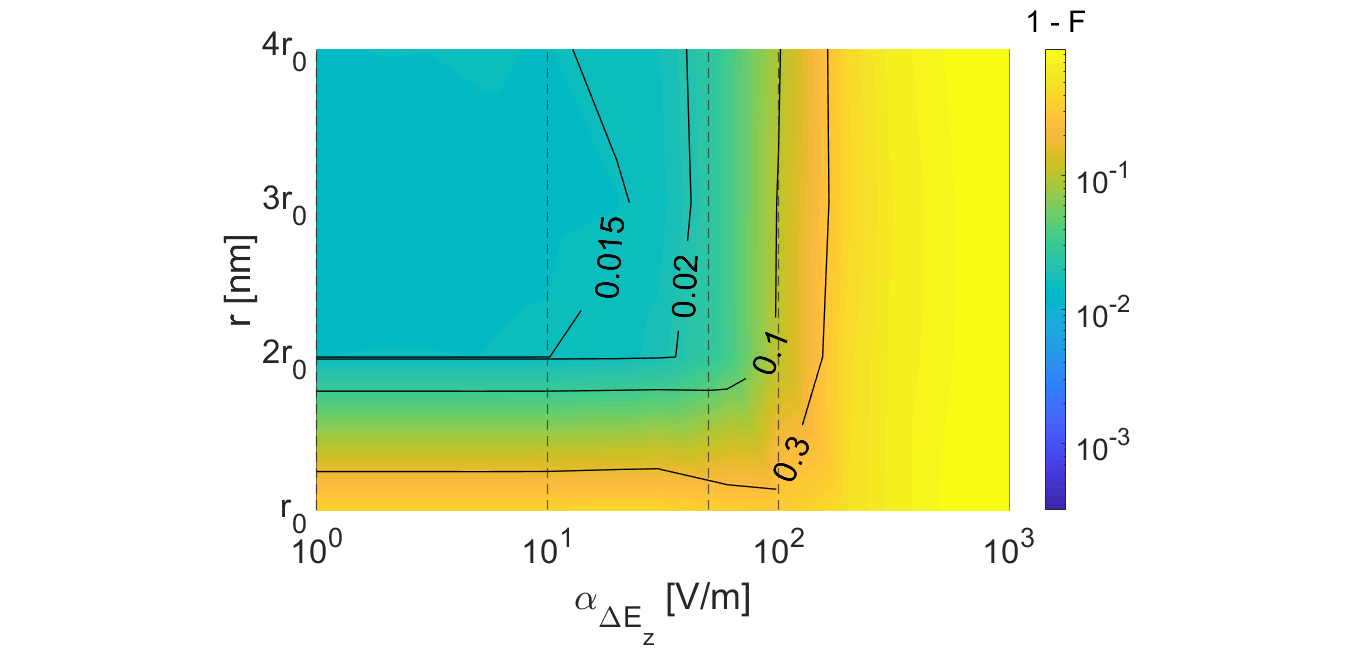}
    \caption{Entanglement infidelity for the operation $R_x ( -\frac{\pi}{2} ) \otimes R_x ( -\frac{\pi}{2} )$ as a function of noise amplitude $\alpha_{\Delta E_z}$ and $r$. Like in $R_z ( -\frac{\pi}{2} ) \otimes R_z (-\frac{\pi}{2} )$ case, the infidelity is deteriorated when $r = r_0$ and for high values of $\alpha_{\Delta E_z}$. The dashed vertical lines highlight the $\alpha_{\Delta E_z}$ values investigated in the next Figure.}
    \label{Fig:Inf-alpha-rcouple_ParRx-0.5pi}
\end{figure}

Similarly to the case of two parallel $\hat{z}$-axis rotations, the infidelities of two parallel $R_x(-\frac{\pi}{2})$ gates are studied. The results of \textbf{Figure \ref{Fig:Inf-alpha-rcouple_ParRx-0.5pi}} show the 1-F results as a function of the inter-qubit distance $r$ and of the noise amplitude $\alpha_{\Delta E_z}$. 

As seen in Figure \ref{Fig:Inf-alpha-rcouple_ParRz-0.5pi}, the parallel operations are compromised by the long-range dipole-dipole interaction when $r = r_0$ while infidelity is kept low for $r \geq 2 r_0$. In this region 1-F is more affected by the noise and its amplitude increase can raise 1-F up to exceed $10^{-1}$ when $\alpha_{\Delta E_z} = 100$ V/m.
Note that the values of the $R_x ( -\frac{\pi}{2} ) \otimes R_x ( -\frac{\pi}{2} )$ infidelity plateaus are higher than those of $R_z ( -\frac{\pi}{2} ) \otimes R_z ( -\frac{\pi}{2} )$ for the same $\alpha_{\Delta E_z}$.

\begin{figure}[htbp!]
    \centering
    \includegraphics[width=0.8\textwidth]{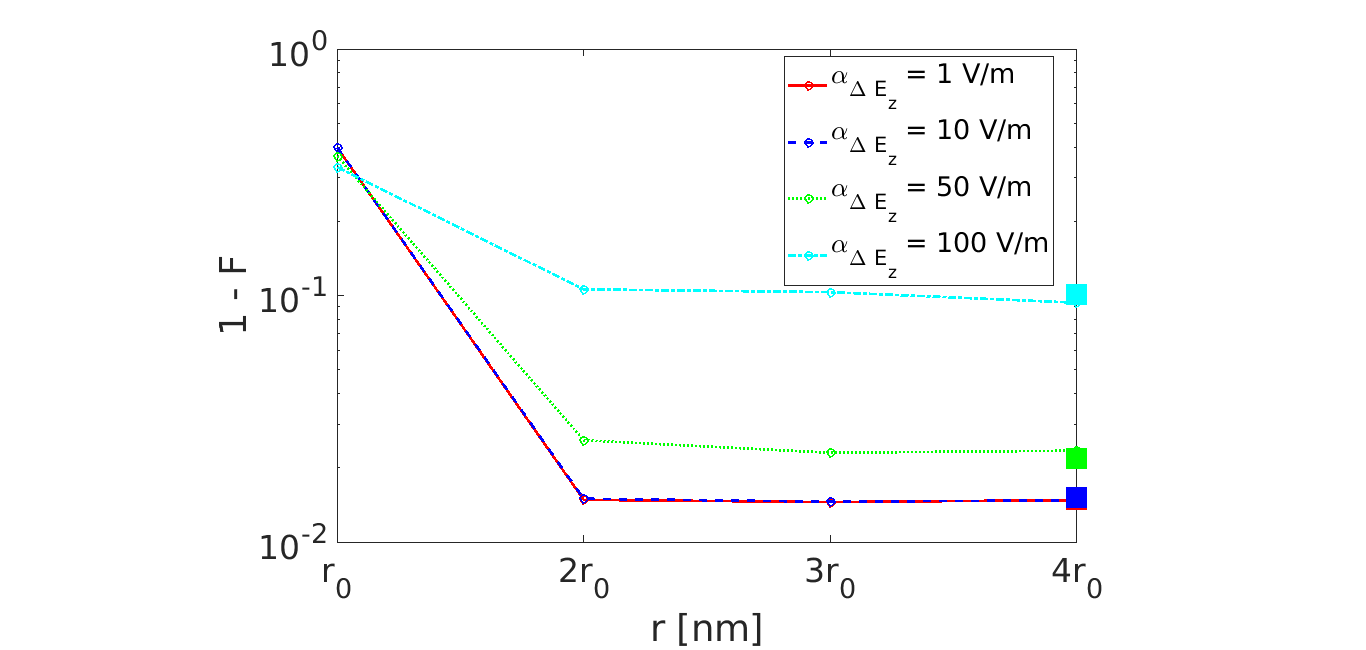}
    \caption{Entanglement infidelity for the operation $R_x ( -\frac{\pi}{2} ) \otimes R_x ( -\frac{\pi}{2} )$ calculated as a function of $r$ for different noise amplitudes $\alpha_{\Delta E_z}$. Except for $r < 2 r_0$, the curves are predominantly influenced by $\alpha_{\Delta E_z}$. The squares represent the infidelity of two non-interacting qubits for each value of $\alpha_{\Delta E_z}$.}
    \label{Fig:Inf-r_ParRx-0.5pi}
\end{figure}

Then, the effect of the inter-qubits distance $r$ on the infidelity of two parallel $R_x ( -\frac{\pi}{2} )$ gates is illustrated in \textbf{Figure \ref{Fig:Inf-r_ParRx-0.5pi}} for four different values of $\alpha_{\Delta E_z}$. Like in Figure \ref{Fig:Inf-r_ParRz-0.5pi}, the infidelities reach their maximum at $r = r_0$, while for $r \geq 2 r_0$ the infidelities are only deteriorated by the increase of $\alpha_{\Delta E_z}$. The fidelity reaches 99\% in correspondence to $\alpha_{\Delta E_z}\leq$ 10 V/m and drops up to 90\% for $\alpha_{\Delta E_z}$=100 V/m.

\section{Parallel two-qubit gate: $\sqrt{iSWAP}$}

\label{Sec:ParallelTwoQubitOp}

\begin{figure}[htbp!]
    \centering
    \includegraphics[width=0.5\textwidth]{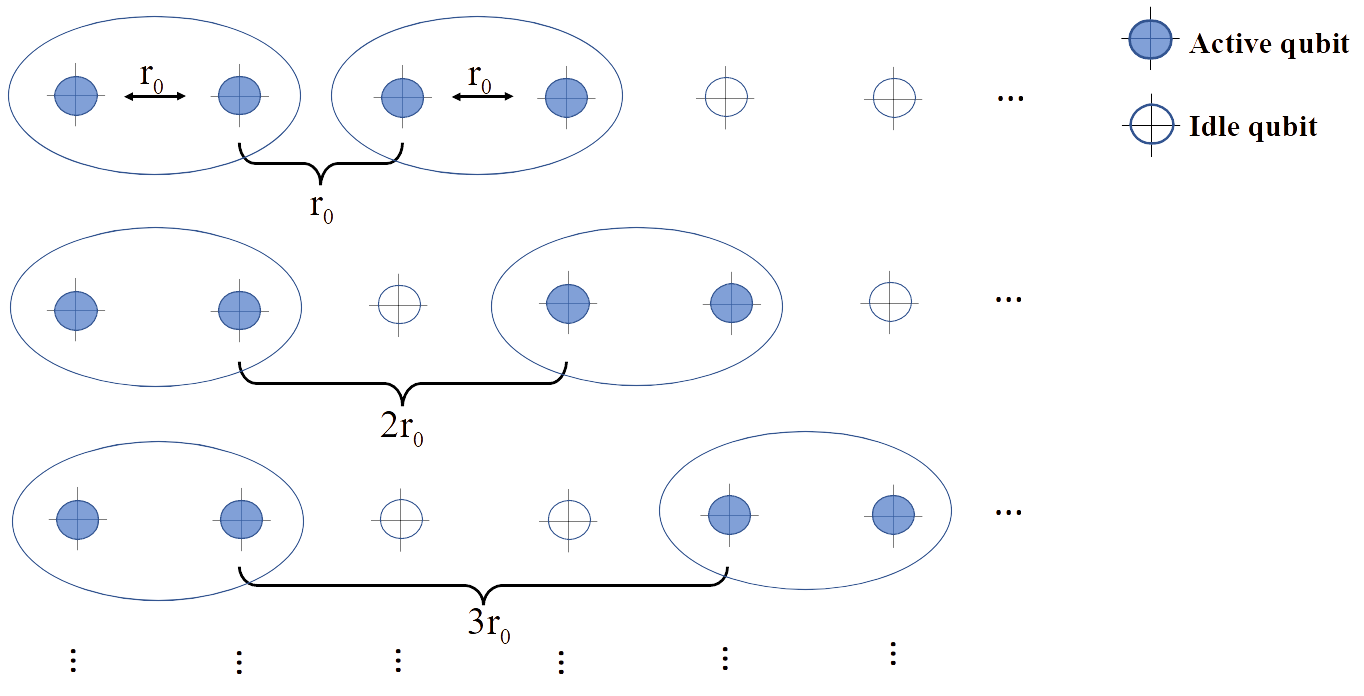}
    \caption{Scheme of a linear array of flip-flop qubits for three example cases when two parallel two-qubit gates are considered. The distance between the qubits operated by two-qubit gate is $r_0$.  
    The first couple of qubits is separated from the second couple by a distance $r$, integer multiple of $r_0$.}
    \label{Fig:ParallelTwoQubitOp}
\end{figure}

In this section the parallel application of two two-qubit operations are studied following the scheme shown in \textbf{Figure \ref{Fig:ParallelTwoQubitOp}}.

A $\sqrt{iSWAP}$ gate on a qubit couple is achieved between two donors spaced by $r_{0} = 180$ nm. First, a DC electric field is applied to both qubits $q_i$ and $q_j$ and finally two identical corrective single $R_z$ gates manipulate the qubits one by one, while the other is kept in an idling state \cite{Ferraro-2021}. All the parameters are reported in Table \ref{tab:SQRTiSWAP}. 

\begin{center}
\begin{table}[htbp!]
  \begin{center}
    \caption{Single $\sqrt{i \, SWAP}$ gate parameters.}
    \begin{tabular}{@{}lllllllll@{}}
    \hline
 & $V_t$ & $\Delta E_{idle}$ & $\Delta E_{int}$ & $\Delta E_{ct}$ & $\tau_1$ & $\tau_2$ & T &  K \\
 &[GHz] &\textrm{[V/m]} & [V/m] & [V/m] & [ns] & [ns] & [ns] &  \text{} \\
      \hline
$q_i q_j$ &11.58 &10000 & 1300 & 0  & 1.3 & 195 & 2 &  $\simeq$ 20 \\
$q_i (q_j)$ &11.58 &10000 & 1300 & 0 & 2 & 4 & 4.5 &  $\simeq$ 33 \\
\hline
\end{tabular}
    \label{tab:SQRTiSWAP}
  \end{center}
\end{table}
\end{center}

In order to study the four-qubit system $i,j,k,l$, its Hamiltonian $\hat{H}_4^{ijkl}$ which, similarly to the two-qubit system, is obtained as the sum of the two-qubit Hamiltonians $\hat{H}_2^{ij}$ (Equation \ref{eq:H2ff}) and the interaction term (Equation \ref{eq:Vdip0}) between only the first nearest neighbours qubits ($j$ and $k$) and neglecting the others, is expressed by
\begin{equation}
\hat{H}_4^{ijkl}=\hat{H}_2^{ij}\otimes\hat{\mathbb{1}}+\hat{\mathbb{1}}\otimes\hat{H}_2^{kl}+\hat{H}_{int}^{jk}.
\end{equation}
 
\begin{figure}[htbp!]
    \centering
    \includegraphics[width=0.8\textwidth]{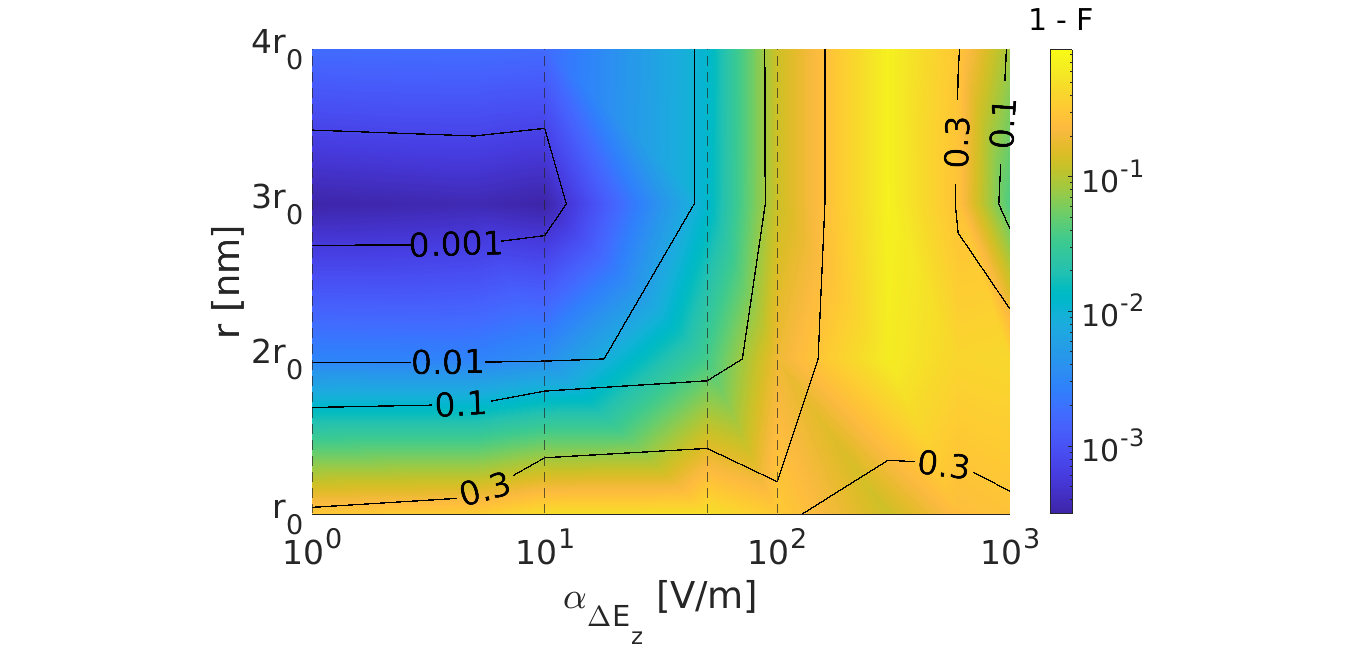}
    \caption{Entanglement infidelity for the operation $\sqrt{i \, SWAP} \otimes \sqrt{i \, SWAP}$ as a function of noise amplitude $\alpha_{\Delta E_z}$ and $r$. The infidelity has its highest values when the dipole-dipole interaction between the qubits $q_2$ and $q_3$ is stronger and where $\alpha_{\Delta E_z}$ is higher. The dashed vertical lines highlight the $\alpha_{\Delta E_z}$ values investigated in the next Figure.}
    \label{Fig:Inf-alpha-rcouple_ParSqrtiSWAP}
\end{figure}

The infidelities trends of two parallel $\sqrt{iSWAP}$ gates are shown in \textbf{Figure \ref{Fig:Inf-alpha-rcouple_ParSqrtiSWAP}} as a function of the distance $r$ and of the noise amplitude $\alpha_{\Delta E_z}$. In particular, similarly to the two parallel one-qubit gates, the infidelity reaches its maximum at $r = r_0$, while for $r \geq 2 r_0$, the noise amplitude dominates the behaviour of 1-F, which exceeds $10^{-1}$ when $\alpha_{\Delta E_z} > 100$ V/m.

\begin{figure}[htbp!]
    \centering
    \includegraphics[width=0.8\textwidth]{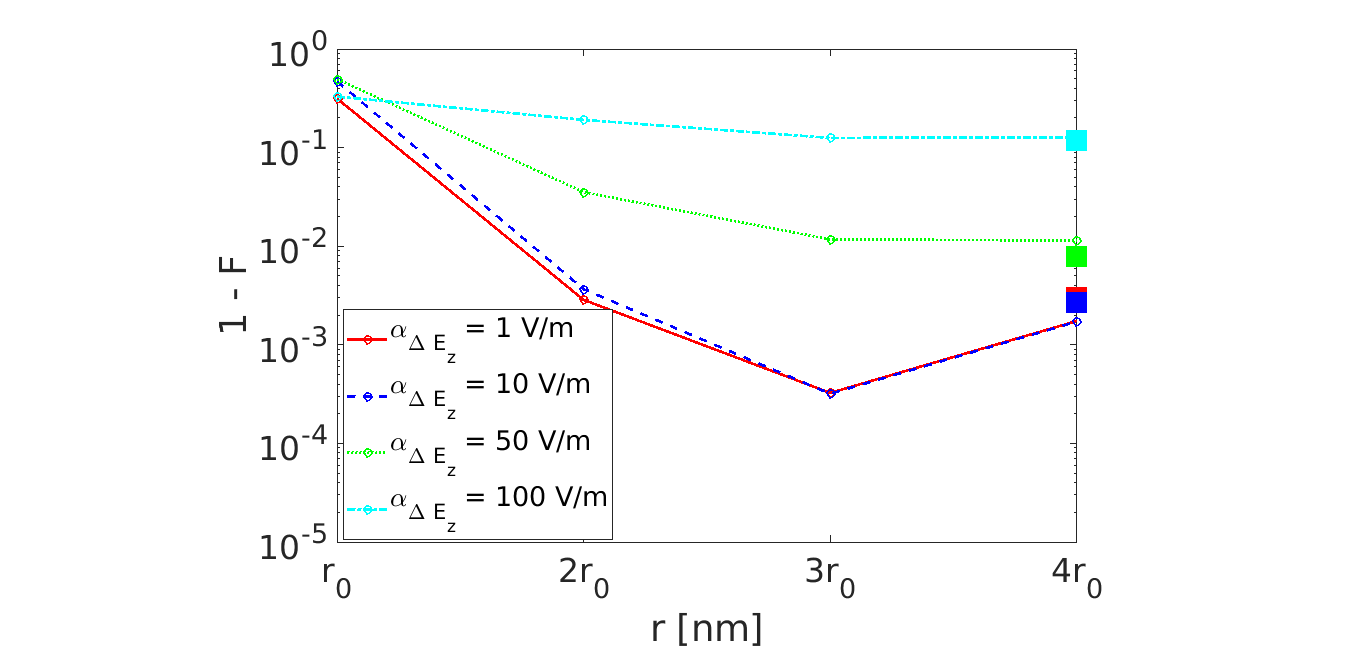}
    \caption{Entanglement infidelity for the operation $\sqrt{i \, SWAP} \otimes \sqrt{i \, SWAP}$ calculated as a function of $r$ for different noise amplitudes $\alpha_{\Delta E_z}$. Similarly to the two parallel one-qubit gates, when $r \geq 2 r_0$, the curves are predominantly influenced by $\alpha_{\Delta E_z}$. The squares represent the infidelity of two non-interacting couples of qubits for each value of $\alpha_{\Delta E_z}$.}
    \label{Fig:Inf-rcouple_ParSqrtiSWAP}
\end{figure}

Finally, the effect of the qubit distance $r$ on 1-F is presented in \textbf{Figure \ref{Fig:Inf-rcouple_ParSqrtiSWAP}} for different values of $\alpha_{\Delta E_z}$. The infidelities reach their maximum values at $r = r_0$ where the dipole-dipole interaction between the second and third qubits has the highest impact, while for $r \geq 2 r_0$  the 1-F increases only by a $\alpha_{\Delta E_z}$ rise due to a negligible interaction between the qubit couples. The fidelity varies from a maximum value of 99.9\% to 90\% when $\alpha_{\Delta E_z}$ increases. In particular a curve minimum in $r = 3r_0$ can be observed when $\alpha_{\Delta E_z} = 1, 10$ V/m whereas no minimum results for higher noise levels. 

The presence of an infidelity minimum for small $\alpha_{\Delta E_z}$ could be explained as a favorable coupling effect between remote couples of qubits. It is worth recalling that each couple of qubits is manipulated with a $\sqrt{iSWAP}$ sequence that has been optimized to control a single couple of qubits thus no direct conclusions on a monotonic decreasing behaviour of the infidelity of two parallel operations as a function of $r$ can be inferred. The simulation results show that for $r> 3 r_0$ the infidelity of the parallel operations raises, approaching that one of two non-interacting couples of qubits (represented by the squares in Figure \ref{Fig:Inf-rcouple_ParSqrtiSWAP}) because the dipole-dipole interaction between the qubit couples is almost negligible for large $r$. For $r< 3r_0$ the two couples of qubits are too close and their unwanted coupling is strongly detrimental for the fidelity of the parallel gates.

\section{Conclusion}

\label{Sec:Conclusion}

In this work the infidelity of two parallel $R_{z}$, $R_{x}$ and $\sqrt{iSWAP}$ gates applied on flip-flop qubits arranged in a linear array are studied. The detrimental effects on the entanglement gate infidelity due to the mutual qubit interference and to the 1/f noise are taken into account. The results obtained show a greater robustness of $R_z ( -\frac{\pi}{2} ) \otimes R_z ( -\frac{\pi}{2} )$ to the noise with respect to the other two parallel operations.

Moreover, a minimum inter-qubit distance $r_{min}=2 r_0$ for each considered gate can identify a safe region for $r \geq r_{min}$ where high-fidelity parallel gates can be achieved. In this safe region, the three operations considered in this study show a good robustness to the 1/f noise until a noise amplitude of 50 V/m, with corresponding infidelities roughly below $10^{-1}$.

It is recognized that is fundamental for quantum computation not only a small gate time with respect to the qubit coherence time but also a sufficiently high level of parallelism. As a first approximation, the results obtained for two parallel operations on two qubits (two couples of qubits) can be extended to an arbitrary number of parallel operated qubits (couples of qubits). 
The resulting $r_{min}= 2 r_0$ implies that at least an idle qubit 
is needed between two active ones to retain high-fidelity parallel one-qubit gates. Therefore, one-qubit gate transversality cannot be obtained in a single time step, with all the gates executed in parallel. Nevertheless, one-qubit gate transversality can be anyway achieved with a gate serialization in two consecutive steps: in the first one, qubits with odd (even) array indexes are operated in parallel and then, in the second step, the even (odd) indexed qubits are manipulated in parallel. A similar consideration can be done for the serialization of parallel two-qubit gates: in the first time step, a set of qubit couples selected in such a way to leave an idle couple between the two active ones are operated and then, during the second time step, the qubit couples that were idling during the first step are manipulated and the remaining ones are left in idle. Surely this leads to a one- (two-) logical qubit gate time that are two times longer than the one- (two-) qubit gate duration but it represents a useful scheme that allows the parallelization of quantum gate operations in view of the realization and exploitation of quantum error correction circuits.

\bibliographystyle{unsrt}
\bibliography{Ref}

\end{document}